\documentclass[twocolumn]{article}

\pdfoutput=1

\usepackage{graphicx}
\usepackage{setspace}
\usepackage[left=2.5cm, right=2.5cm, top=2.5cm, bottom=2.5cm]{geometry}
\usepackage{amsmath}
\usepackage{amssymb}
\usepackage{svg}
\usepackage{wrapfig}
\usepackage{subcaption}
\usepackage{adjustbox}
\usepackage{graphicx}
\usepackage{authblk}
\usepackage{gensymb}
\usepackage{abstract}
\usepackage{enumitem}
\usepackage{hyperref}
\usepackage{textcomp}
\usepackage{siunitx}

\title{\textbf{Redundant Cross-Correlation for Drift Correction in SEM Nanoparticle Imaging}}
\date{October 30, 2024}

\author[1,2]{Bischoff Montenegro, Iago}
\author[1]{Prikoszovich, Konrad}
\author[1]{Lee, Subin}
\author[1,2]{Quiring, Kilian}
\author[3]{Zimmerman, Jonathan}
\author[1]{Kirchlechner, Christoph}

\affil[1]{\footnotesize Institute for Applied Materials - Mechanics of Materials and Interfaces, Karlsruhe Institute of Technology, Karlsruhe, Germany}
\affil[2]{\footnotesize Hector-Seminar Standort Karlsruhe, Karslruhe, Germany}
\affil[3]{\footnotesize Department of Materials Science and Engineering, Technion, Israel Institute of Technology, Haifa, Israel}

\begin{document}

\setlist[description]{leftmargin=1em}

\twocolumn[
\begin{center}
  \maketitle
\end{center}

\begin{onecolabstract}
  \setlength{\leftskip}{0.5cm}
  \setlength{\rightskip}{0.5cm}
  \noindent
  Scanning Electron Microscopy (SEM) is a widely used tool for nanoparticle characterization, but long-term directional drift can compromise image quality. We present a novel algorithm for post-imaging drift correction in SEM nanoparticle imaging. Our approach combines multiple rapidly acquired, noisy images to produce a single high-quality overlay through redundant cross-correlation, preventing drift-induced distortions. The preservation of critical geometrical properties and accurate imaging of surface features were verified using Atomic Force Microscopy. On platinum nanoparticles with diameters of 300 to 1000 nm, significant improvements in the mean-based signal-to-noise ratio (SNR) were achieved, increasing from 4.4 dB in single images to 11.3 dB when overlaying five images. This method offers a valuable tool for enhancing SEM image quality in nanoparticle research and metrology, particularly in settings without specialized hardware-based drift correction.
\end{onecolabstract}
\vspace{1 cm}
]

\section{Introduction}

Scanning electron microscopy (SEM) is a widely used imaging technique for nanoparticles \cite{vladar2020overview}. However, it is prone to scan artifacts such as vibrations and scan drift. Long-term directional drift, often caused by sample charging or thermal drift, results in blur and image corruption \cite{cizmar2011}, making it problematic for applications requiring high-precision metrology.

\begin{figure}[ht]
    \centering
    \includegraphics[width = 0.45\textwidth]{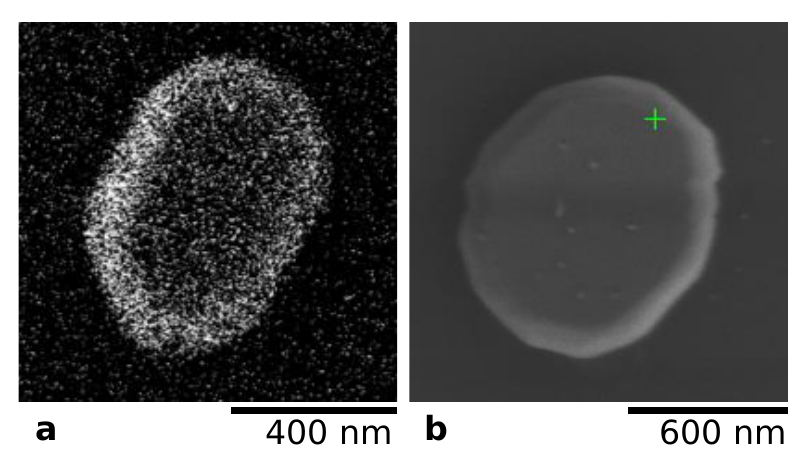}
    \caption{Fast scan speed image with high noise levels (a) and slow scan speed image with distorsions (b)}
    \label{fig:fast-vs-slow}
\end{figure}

While fast scan speeds can minimize long-term directional drift, they introduce high noise levels that obscure sample details \cite{10284219, Oho2020} (see Fig. \ref{fig:fast-vs-slow}). Overlaying multiple noisy images can reduce noise, producing a higher-quality image without translational artifacts.

This process requires compensating for sample drift during the acquisition of an image series. Several algorithms have been proposed to address this issue. The Lucas-Kanade algorithm \cite{lucas1981} estimates optical flow by determining spatial and temporal intensity gradients at different points in the field of view \cite{yao2023}. However, it assumes consistent brightness across frames \cite{lucas1981}, limiting its effectiveness in noisy environments.

Cross-correlation offers a more noise-tolerant alternative \cite{mantooth2002}, searching for the correct alignment between two images by measuring their similarity for different shifts. Although it is less accurate than the Lucas-Kanade algorithm \cite{yao2023}, it is more resilient against noise \cite{evangelidis2007photometric}, particularly when using redundant cross-correlation \cite{wang2014}. This makes it ideal for image series with low pixel dwell times. However, many cross-correlation techniques, such as those in \cite{mantooth2002} and \cite{snella2010drift}, are complex, often involving Fourier transforms. This paper proposes a simpler, direct cross-correlation method with minimal pre-processing, suitable for rapid image acquisition and analysis.

\section{Materials and Methods}

\subsection{Image acquisition}

Platinum nanoparticles with diameters of roughly \(300-1000 \text{nm}\) on a sapphire substrate were used. The nanoparticles were produced via solid-state dewetting \cite{dupraz2022imaging}. Image series containing up to 68 images, were acquired with a ZEISS Merlin Gemini II microscope operated at an acceleration voltage of \(2kV\) and a beam current of \(200 pA\). Scan speeds of 3 and 4, corresponding to pixel dwell times of 0.435 $\mu s$ and $0.845\mu s$ respectively, were used. The 1024x768 pixel 8bit-greyscale images were taken from the secondary electron detector operated at a bias of \(300 V\).

Prior to cross-correlation, the images were preprocessed with a Gaussian blur to mitigate the influence of single-pixel noise. A 13×13 kernel provided the best performance.

\subsection{Algorithm}

The aim of this algorithm is to convert a sequential image series \(P\) into a single overlay image. To achieve this, it is essential to determine the correct shift vector \(v^*_{ij}\ \in \mathbb{Z}^2\) between any two images \(p_i, p_j \in P\). This is done by evaluating a cost function for different possible shift vectors \(v_{ij}\).

\subsubsection{Cost function}

The cost function \(f: v_{ij} \mapsto c\)  measures the average differences in pixel values between two images shifted by a certain vector \(v_{ij}\ \in \mathbb{Z}^2\). It is defined as

\[f(v_{ij}) = \sum_{a,b \in Q,R} (q_{ab} - r_{ab})^2 \cdot \frac{1}{|Q|}\]

where \(Q\) and \(R\) represent the remaining common parts of \(p_i\) and \(p_j\) after shifting by \(v_{ij}\), and \(|Q|\) (which is equal to \(|R|\)) denotes the total number of pixels in this common part (excluding a region near the image edges to avoid comparability issues). The terms \(q_{ab}\) and \(r_{ab}\) are the individual pixel values from \(Q\) and \(R\) respectively at the \(x\) and \(y\) coordinates \(a\) and \(b\). This function was chosen due to its low computational complexity and good performance on the tested image series.

\subsubsection{Minimum search}

\(f(v_{ij})\) is minimal when \(v_{ij} = v^*_{ij}\), i.e. when the images are correctly overlaid. In this case, each pixel in \(p_j\) corresponds to the pixel in \(p_i\) that represents the same part of the sample, with any remaining differences attributed solely to background noise. As \(v_{ij}\) deviates from \(v^*_{ij}\), this correspondence is lost, the pixel value difference substantially increases, particularly near boundaries between different features (e.g. the background and the nanoparticle), raising the average \(f(v_{ij})\).

\begin{figure}[ht]
    \centering
    \includegraphics[width=0.45\textwidth]{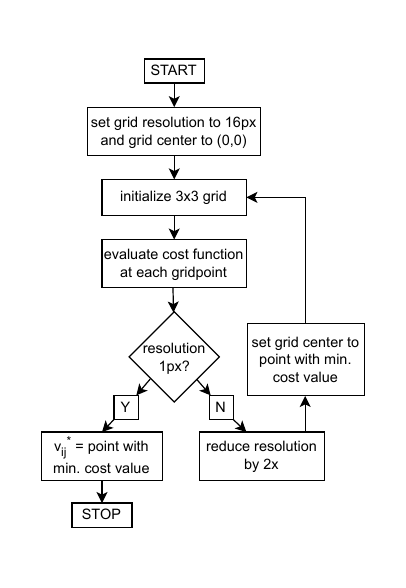}
    \caption{Grid search algorithm}
    \label{fig:gridsearch_flowchart}
\end{figure}

The minimum of the cost function is determined using a multi-scale grid search process, as detailed in Fig \ref{fig:gridsearch_flowchart}. The process typically begins with a relatively coarse grid spacing (16 pixels in this study) and iteratively refines it using a 3×3 grid to evaluate the cost function at each step. This refinement continues until the grid resolution reaches 1 pixel, at which point the algorithm determiness \(v^*_{ij}\) as the shift vector that minimizes the cost function. Parameters such as the starting resolution can be adjusted to meet the requirements of any specific image series.

\subsubsection{Redundant cross-correlation}

This approach can lead to minor errors in vector calculation, as the impact of background noise remains significant near the cost function's minimum. To mitigate this, the algorithm incorporates a redundant cross-correlation method. This method not only considers the direct vector \(v_{ij}\) but also examines vectors \(v_{jk} - v_{ik}\) for all other images \(k\) in the series, resulting in \(n-2\) additional vectors for a series of \(n\) images. The final shift vector \(v^*_{ij}\)  is determined by taking the median of the x and y coordinates of all these vectors, thereby reducing the influence of random noise and eliminating outliers. This method is similar to the drift correction technique described by \cite{wang2014}, originally developed for super-resolution localization microscopy, but is significantly simpler.

If all possible shifts are calculated, the computational complexity of this algorithm is \(O(N^2P)\), with \(N\) being the number of images in the series and \(P\) the number of pixels in each of them. Since this scales rapidly for large datasets (see Discussion), a function was implemented that calculates the shifts of all images relative to a fixed set of reference images instead of all images. This modification reduces the computational complexity to \(O(NP)\) without compromising overlay image quality, provided the number of reference images \(k\) is sufficiently large.

\subsection{Verification experiments}
\label{sec:afm-verification}

Atomic force microscopy (AFM) was used to verify the shape and surface details of nanoparticles in a correlative fashion \cite{Bellotti2022}\cite{Rao2007}\cite{Zimmerman2021} . The measurements were performed on a Bruker Dimension Icon 4000 at the Institute of Microstructure Technology (IMT) at KIT using monolithic silicone cantilevers with a reflective aluminum coating (BudgetSensors, All-In-One-Al, Type C)  and a tip radius of 10 nm. To validate our algorithm we both use the Height Sensor data as well as the Amplitude Error data. The image acquisition parameters for the different particles are shown in  Tab \ref{tab:AFM_Img_param}. Particle 1 was investigated using the PeakForce tapping mode \cite{Xu2018} \cite{Bairamukov2022} of the Dimension Icon due to movement of the particle during normal tapping mode. This results in the different values in lines and samples per line for this particle.

\begin{table}[ht]
    \centering
    \begin{tabular}{|c|c|c|c|} \hline 
         Part. &  Lines&  Samples/Line& Img. dimensions\\ \hline 
         1&  400&  400& 8$\mu  m$ x 8$\mu  m$\\ \hline
         2&  346&  608& 8 $\mu  m$ x 8$\mu  m$\\ \hline
    \end{tabular}
    \caption{AFM image parameters for each particle.}
    \label{tab:AFM_Img_param}
\end{table}

\section{Results}

\begin{figure*}[ht]
    \centering
    \includegraphics[width=0.95\textwidth]{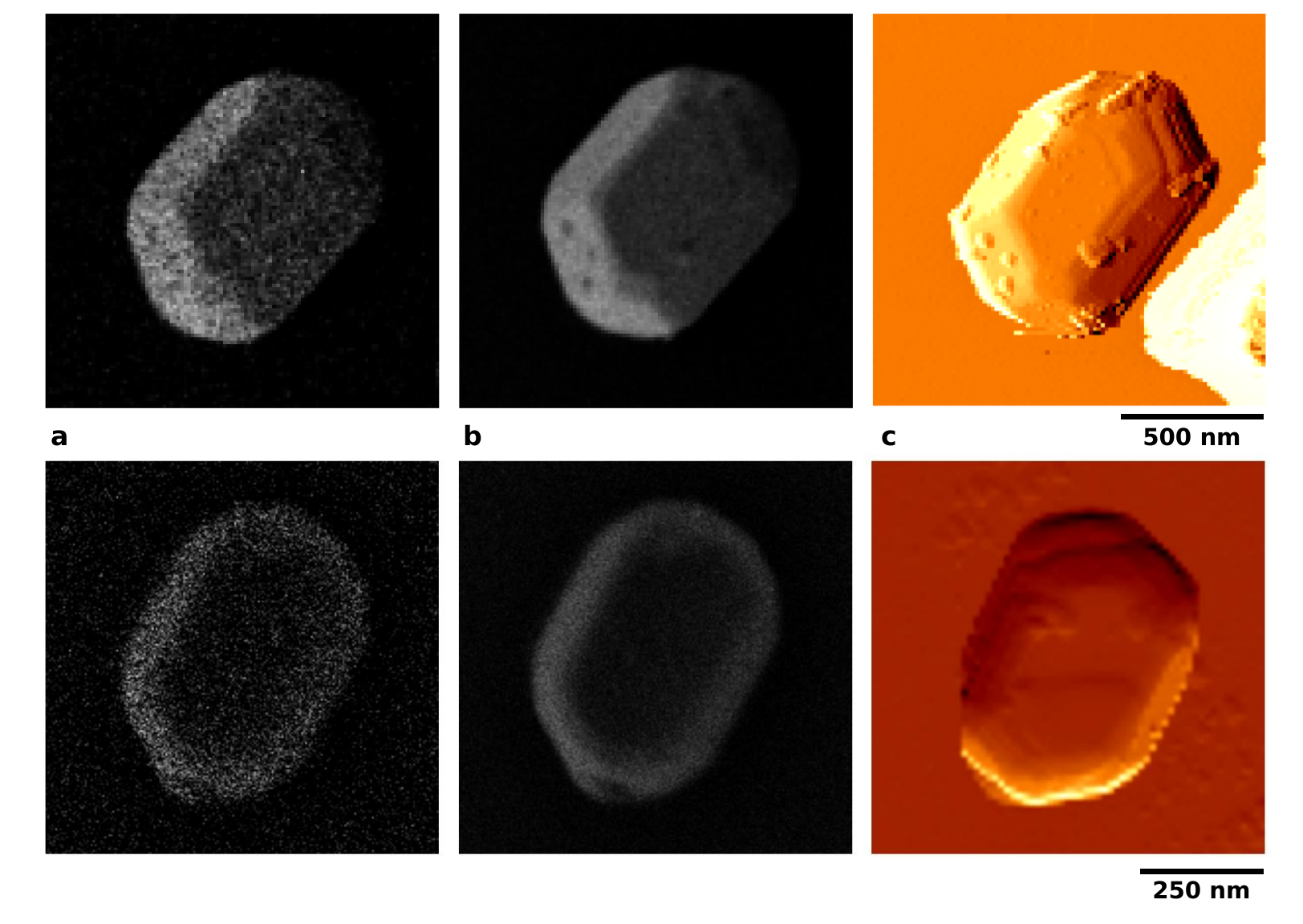}
    \caption{Comparison of a raw SEM image (a), the overlay image produced by our algorithm using 14 raw images (b) and the amplitude error AFM plot corresponding to the same particle (c) for two different nanoparticles, referred to as particle 1 (top) and particle 2 (bottom) in the text. The bright area at the bottom right of particle 1's AFM plot is an imaging artifact caused by a nearby nanoparticle.}
    \label{fig:3_comp}
\end{figure*}

Figure \ref{fig:3_comp} shows the results of our algorithm applied to two distinct nanoparticles, comparing a single raw SEM image, an overlay image created from 14 SEM images using our algorithm, and an AFM image of the same particle.

\begin{figure}[ht]
    \includegraphics[width=0.45\textwidth]{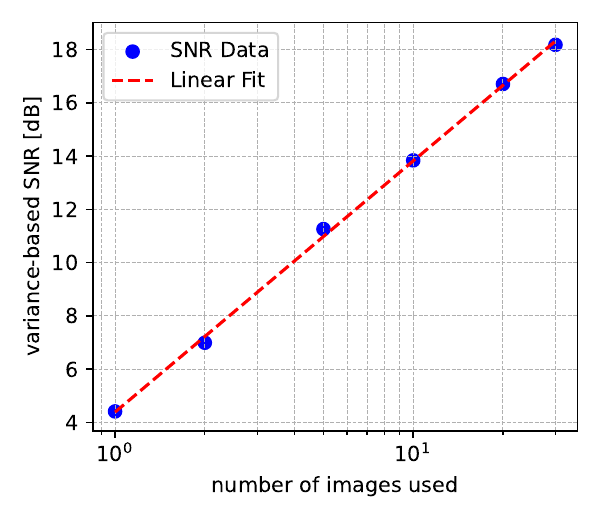}
    \caption{SNR for overlay images with different numbers of raw images. A logarithmic fit \(4.1 \log{x} + 4.38\) with \(R^2=0.9989\) is also included. These values have a \(\pm O(0.1)\) error margin, estimated by taking the SNR in several parts of the image.}
    \label{fig:snr-improve}
\end{figure}

\subsection{Overlay image vs raw SEM}

The overlay image demonstrates a significant reduction in noise compared to the raw SEM image. A mean-based signal-to-noise ratio (SNR) analysis was conducted using a 400-pixel area on the top facet of particle 1, where uniform brightness is assumed. Mean-based SNR is defined as

\[\text{SNR} = \frac{\mu^2}{\sigma^2}\]

where \(\mu\) is the average pixel brightness and \(\sigma^2\) is the variance in pixel brightness. This analysis showed that the SNR increases logarithmically with the number of images used for the overlay (Fig. \ref{fig:snr-improve}). The SNR is 4.4 dB for a single image but rises to 11.3 dB when 5 images are overlaid. This increase in SNR not only enhances the clarity of the particle's edges but also improves the visibility of surface features, such as the three dark grains on the bottom left of particle 1 and the shape irregularities at the top and bottom of particle 2.

\subsection{Overlay image vs AFM plot}

The AFM plots provide more detailed surface information and offer a way to independently verify the particles' dimensions in the SEM images. A comparison of the ratio between the two major axes of each particle shows a \(-0.1 \%\) discrepancy for particle 1 and a \(+0.7 \%\) discrepancy for particle 2. Angle measurements between the two inclined bottom sides yield \(60\degree \pm 1\) for the SEM overlay and \(59.7\degree \pm  0.7\) for the AFM, indicating good angle preservation in the SEM image. It is also possible that the AFM plots themselves are distorted. Not only did sample movement cause problems during the data acquisition (see Section \ref{sec:afm-verification}), they also exhibit several artifacts, such as the bright area to the right of particle 1 or the straight line on particle 2's bottom left edge, likely caused by parts of the cantilever further away from the tip coming in contact with a larger nanoparticle close to the investigated particles.

\section{Discussion}

\subsection{Analysis of results}

The results indicate that our algorithm effectively preserves the geometric properties of the imaged nanoparticles. The difference between the angle measurements in the SEM overlay image and the AFM is within the margin of error, and while the \(0.7 \%\) discrepancy in the major axis ratio for particle 2 is notable, it remains unclear whether this discrepancy originated from our SEM overlay image or from artifacts in the AFM imaging process. Further verification experiments, ideally employing different microscopy methods, may be necessary.

The results from SNR analysis follow the expected logarithmic pattern: Each additional image in the average reduces the noise by a constant factor \(r\) while the signal remains constant. This increases SNR by a factor of \(\sqrt{r}\), corresponding to logarithmic improvement on the dB scale. This enhancement is substantial, and with a sufficient number of SEM images, features that are lost in the noise in a single frame can be clearly identified and interpreted. An example are the black regions on the bottom left of particle 1 (Fig. \ref{fig:3_comp}).

\subsection{Limitations}

Despite its accuracy and resilience to noise, our algorithm has certain limitations that should be considered:

\begin{description}
    \item[Feature distinction] The cost function requires a clear difference in greyscale values between the image background and features that are about as large typical drift between any two images. Otherwise, there might not be a clear gradient towards a global minimum, leading to errors in shift vector estimation.
    \item[Drift magnitude] The maximum drift between any two images in the series should not be excessive. Besides increasing starting grid size and therefore computation time, this reduces the common image area in some drift calculations. With a lower total number of pixels, the influence of remaining random noise on the average difference increases, making the cost function less reliable.
    \item[Computational demands] Runtime tests were conducted using our 1024x768 pixel images on a consumer-grade Windows laptop (AMD Ryzen 5 5500U, 16 GB of RAM). While processing times for smaller image sets were negligible (1.6s for 5 images, 6.5s for 10 images), larger series require significantly more time due to the \(O(N^2P)\) computational complexity.
    
    The use of partial graph calculations mitigates this issue (24.8s for 30 images with \(k=5\)) and could be tested with larger series containing \(\mathcal{O}(100-1000)\) images in the future. Improvements could also be achieved by implementing the algorithm using a programming language with faster array operations, e.g. Mojo.
\end{description}

\section{Conclusion}

We present an algorithm that creates high-quality SEM images of nanoparticles by combining multiple low-quality, drift-free images using redundant cross-correlation. The algorithm employs a cost function and multi-step grid search to determine drift between images. Key achievements include:

\begin{enumerate}
    \item Mean-based SNR increased from 4.4 dB (single image) to 11.3 dB (5-image overlay), following a logarithmic trend.
    \item Preservation of particle shape and size, verified by Atomic Force Microscopy.
\end{enumerate}

Our algorithm is valuable for both high-precision metrology and qualitative studies requiring size and angle preservation, enabling measurements on SEMs without built-in drift correction. However, limitations include dependence on distinguishable image features, sensitivity to excessive drift variation, and high computational demands. Future work could optimize the computational efficiency. Additionally, verification using other microscopy techniques would further confirm size and angle preservation.

\section*{Acknowledgments}

The authors wish to express their gratitude towards the Hector-Seminar sponsered by the H.W. \& J. Hector Foundation zu Weinheim for its invaluable support in the first phase of this project. Special thanks go to Mr. Norbert Krieg for his guidance, and to Dr. Hans-Werner and Josephine Hector, whose generous support of the Hector foundation made this work possible. We also acknowledge support of the Karlsruhe Nano Micro Facility (KNMF), a Helmholtz Research Infrastructure at Karlsruhe Institute of Technology (KIT) for instrument time at the AFM, as well as R. Thelen and C. F. Pichler (IMT KIT) for their assistance with AFM measurements and data readout scripts. Financial support from KIT Campus Transfer GmbH during the second funding phase of this project is also acknowledged.

\section*{Supplementary Materials}
\label{sec:supplementary-materials}

The full implementation of the algorithm in Python is available at \href{https://github.com/iago-bm/NanoPEACH_code}{https://github.com/iago-bm/NanoPEACH\_code}.

\begingroup
    \footnotesize
    \bibliographystyle{unsrt}
    \bibliography{references}
\endgroup

\end{document}